\documentclass[12pt,onecolumn]{IEEEtran}
\usepackage{amsmath}
\usepackage{amssymb}
\usepackage{amsfonts}
\usepackage{graphicx}

\allowdisplaybreaks
\input{tcilatex}

\begin{document}

\begin{center}
\bigskip {\huge \ Two-step differentiator for delayed signal}

Xinhua Wang and Hai Lin

{\small Department of Electrical \& Computer Engineering, National
University of Singapore, Singapore 117576}

{\small Email: wangxinhua04@gmail.com Tel: 65 97756061}
\end{center}

\emph{Abstract:} This paper presents a high-order differentiator for delayed
measurement signal. The proposed differentiator not only can correct the
delay in signal, but aslo can estimate the undelayed derivatives. The
differentiator consists of two-step algorithms with the delayed time
instant. Conditions are given ensuring convergence of the estimation error
for the given delay in the signals. The merits of method include its simple
implementation and interesting application. Numerical simulations illustrate
the effectiveness of the proposed differentiator.


{\Large 1. Introduction}

This paper focuses on the problem of estimating derivatives for delayed
signals. Usually measurement and communication of signals all exist time
delays. Obtaining the velocities of tracked targets is crucial for several
kinds of systems with correct and timely performances, such as the
missile-interception systems [1] and underwater vehicle systems [2],
However, the delay phenomenon make it difficult to obtain the undelayed
derivatives of signal. In recent years, researchers tried to apply alternate
methodologies to design differentiators [3-12]. The popular high-gain
differentiators [6, 7, 8] provide for an exact derivative when their gains
tend to infinity. In [9, 10], a differentiator via second-order (or
high-order) sliding modes algorithm was proposed. The information one needs
to know on the signal is an upper bound for Lipschitz constant of the
derivative of the signal. In [11, 12], a finite-time-convergent
differentiator based on finite-time stability and singular perturbation
technique was presented. In all of the aforementioned papers, the signals
are required to be undelayed, otherwise, delayed phenomena happen in
derivative outputs.

This paper provides a design of high-order differentiator for delayed
signal. The proposed differentiator algorithm has a two-step structure. The
differentiator algorithm is composed by two step sub-differentiators, the
first step sub-differentiator estimates the derivatives at the delay, and
the second step sub-differentiator estimates the present derivatives. It is
shown that, under suitable conditions, for a given delay there exists a
differentiator of suitable dimension achieving error decay.



{\Large 2. Problem statement}

For the normal high-gain differentiator [6, 7, 8]:

\begin{eqnarray}
\dot{x}_{i} &=&x_{i+1}+\frac{k_{i}}{\varepsilon ^{i}}\left( v\left( t-\Delta
\right) -x_{1}\right) ,i=1,\cdots ,n-1  \notag \\
\dot{x}_{n} &=&\frac{k_{n}}{\varepsilon ^{n}}\left( v\left( t-\Delta \right)
-x_{1}\right)
\end{eqnarray}%
where $x_{i}$ ($i=1,\cdots ,n$) are the states of differentiator (1), $%
\varepsilon >0$ is the perturbation parameter, and $v\left( t-\Delta \right) 
$ is a signal with time delay $\Delta >0$. The following conclusion can be
obtained:

\begin{equation}
\underset{\varepsilon \rightarrow 0}{\lim }x_{i}=v^{\left( i-1\right)
}\left( t-\Delta \right) ,i=1,\cdots ,n
\end{equation}

From (2), we can find that the delay phenomena happen in the estimation
outputs. And the delay phenomena exist in other differentiators [9-12]. We
are interested in designing a differentiator to force the states of
differentiator to approximate the undelayed derivatives with the delayed
signal $v\left( t-\Delta \right) $.

{\Large 3. Design of high-order two-step differentiator}

Here, we design a differentiator for a delayed signal, and the undelayed
signal tracking and derivatives estimation are obtained from the delayed
signal. The proposed differentiator algorithm has a two-step structure. It
is shown that for a given delay there exists a differentiator of suitable
dimension such that the differentiator states approximate the undelayed
derivatives of signals in spite of the existence of time delay.

We design a two-step high-order differentiator as follow:

\begin{eqnarray}
\dot{x}_{i,1} &=&x_{i+1,1}+\frac{k_{i}}{\varepsilon ^{i}}\left( v\left(
t-\Delta \right) -x_{1,1}\right) ,i=1,\cdots ,n-1  \notag \\
\dot{x}_{n,1} &=&\frac{k_{n}}{\varepsilon ^{n}}\left( v\left( t-\Delta
\right) -x_{1,1}\right)
\end{eqnarray}%
and

\begin{eqnarray}
\dot{x}_{i,2} &=&x_{i+1,2}+\left( \sum\limits_{j{=i}}^{{n}}\frac{1}{\left(
j-i\right) !}\frac{k_{j}}{\varepsilon ^{j}}\Delta ^{j-i}\right) \left(
v\left( t-\Delta \right) -x_{1,1}\right) ,  \notag \\
i &=&1,\cdots ,n-1  \notag \\
\dot{x}_{n,2} &=&\frac{k_{n}}{\varepsilon ^{n}}\left( v\left( t-\Delta
\right) -x_{1,1}\right)
\end{eqnarray}%
where $k_{1}$, $\cdots $, $k_{n}$, are selected such that $%
s^{n}+k_{1}s^{n-1}+\cdots +k_{n-1}s+k_{n}=0$ is Hurwitz, \ and $\varepsilon
>0$ is the perturbation parameter.

\emph{Theorem 1:} For two-step high-order differentiator (3)-(4) and delayed
signal $v\left( t-\Delta \right) $, there exist $k_{1}$,$\cdots $,$k_{n}$,
such that:

\begin{equation}
\underset{\varepsilon \rightarrow 0}{\lim }X_{i,2}\left( s\right)
=s^{i-1}V\left( s\right) ,i=1,\cdots ,n-1
\end{equation}%
approximately. Where $X_{i,2}\left( s\right) $ ($i=1,\cdots ,n-1$) and $%
V\left( s\right) $ are the Laplace transformations of $x_{i,2}\left(
t\right) $, ($i=1,\cdots ,n-1$) and $v\left( t\right) $, respectively.

\emph{Proof:} The laplace transformamtion of (3) is

\begin{eqnarray}
sX_{i,1}\left( s\right) &=&X_{i+1,1}\left( s\right) +\frac{k_{i}}{%
\varepsilon ^{i}}\left( e^{-s\Delta }V\left( s\right) -X_{1,1}\left(
s\right) \right) ,  \notag \\
i &=&1,\cdots ,n-1  \notag \\
sX_{n,1}\left( s\right) &=&\frac{k_{n}}{\varepsilon ^{n}}\left( e^{-s\Delta
}V\left( s\right) -X_{1,1}\left( s\right) \right)
\end{eqnarray}

From (6), we have

\begin{equation}
X_{1,1}\left( s\right) =\left( \sum\limits_{i{=1}}^{{n}}\frac{k_{i}}{%
s^{i}\varepsilon ^{i}}\right) \left( e^{-s\Delta }V\left( s\right)
-X_{1,1}\left( s\right) \right)
\end{equation}

Therefore, we can get

\begin{equation}
X_{1,1}\left( s\right) =\frac{s^{n-1}\varepsilon ^{n-1}k_{1}+\cdots
+s\varepsilon k_{n-1}+k_{n}}{s^{n}\varepsilon ^{n}+\cdots +s\varepsilon
k_{n-1}+k_{n}}e^{-s\Delta }V\left( s\right)
\end{equation}

From (8), we have

\begin{equation}
e^{-s\Delta }V\left( s\right) -X_{1,1}\left( s\right) =\frac{%
s^{n}\varepsilon ^{n}}{s^{n}\varepsilon ^{n}+\cdots +s\varepsilon
k_{n-1}+k_{n}}e^{-s\Delta }V\left( s\right)
\end{equation}

The laplace transformation of (4) is

\begin{eqnarray}
sX_{i,2}\left( s\right) &=&X_{i+1,2}\left( s\right) +\left( \sum\limits_{j{=i%
}}^{{n}}\frac{1}{\left( j-i\right) !}\frac{k_{j}}{\varepsilon ^{j}}\Delta
^{j-i}\right) \left( e^{-s\Delta }V\left( s\right) -X_{1,1}\left( s\right)
\right) ,  \notag \\
i &=&1,\cdots ,n-1  \notag \\
sX_{n,2}\left( s\right) &=&\frac{k_{n}}{\varepsilon ^{n}}\left( e^{-s\Delta
}V\left( s\right) -X_{1,1}\left( s\right) \right)
\end{eqnarray}

From (10), we have

\begin{eqnarray}
X_{i,2}\left( s\right) &=&\sum\limits_{m{=i}}^{{n}}\frac{1}{s^{m-i+1}}\left(
\sum\limits_{j{=m}}^{{n}}\frac{1}{\left( j-m\right) !}\frac{k_{j}}{%
\varepsilon ^{j}}\Delta ^{j-m}\right) \left( e^{-s\Delta }V\left( s\right)
-X_{1,1}\left( s\right) \right)  \notag \\
i &=&1,\cdots ,n
\end{eqnarray}

Therefore, from (11) and (9), we have equation 
\begin{eqnarray}
X_{i,2}\left( s\right) &=&\frac{1}{s^{n-i+1}\varepsilon ^{n}}\sum\limits_{m{%
=i}}^{{n}}s^{n-m}\left( \sum\limits_{j{=m}}^{{n}}\frac{1}{\left( j-m\right) !%
}\varepsilon ^{n-j}k_{j}\Delta ^{j-m}\right) \left( e^{-s\Delta }V\left(
s\right) -X_{1,1}\left( s\right) \right)  \notag \\
&=&\frac{\sum\limits_{m{=i}}^{{n}}s^{n-m}\left( \sum\limits_{j{=m}}^{{n-1}}%
\frac{1}{\left( j-m\right) !}\varepsilon ^{n-j}k_{j}\Delta ^{j-m}\right)
+\sum\limits_{m{=i}}^{{n}}s^{n-m}\frac{1}{\left( n-m\right) !}k_{n}\Delta
^{n-m}}{s^{n-i+1}\varepsilon ^{n}}\left( e^{-s\Delta }V\left( s\right)
-X_{1,1}\left( s\right) \right)  \notag \\
&=&s^{i-1}\frac{\sum\limits_{m{=i}}^{{n}}s^{n-m}\left( \sum\limits_{j{=m}}^{{%
n-1}}\frac{1}{\left( j-m\right) !}\varepsilon ^{n-j}k_{j}\Delta
^{j-m}\right) +k_{n}\left( 1+s\Delta +\cdots +\frac{1}{\left( n-i\right) !}%
s^{n-i}\Delta ^{n-i}\right) }{s^{n}\varepsilon ^{n}}  \notag \\
&&\times \frac{s^{n}\varepsilon ^{n}}{s^{n}\varepsilon
^{n}+s^{n-1}\varepsilon ^{n-1}k_{1}+\cdots +s\varepsilon k_{n-1}+k_{n}}%
e^{-s\Delta }V\left( s\right)
\end{eqnarray}%
for $i=1,\cdots ,n$. Approximately, we can get

\begin{equation}
e^{s\Delta }\approx 1+s\Delta +\cdots +\frac{1}{\left( n-i\right) !}%
s^{n-i}\Delta ^{n-i},i=1,\cdots ,n-1
\end{equation}

Therefore, from (12) and (13), approximately, we have

\begin{equation}
\underset{\varepsilon \rightarrow 0}{\lim }X_{i,2}\left( s\right)
=s^{i-1}V\left( s\right) ,i=1,\cdots ,n-1
\end{equation}%
i.e., we can get that

\begin{equation}
\underset{\varepsilon \rightarrow 0}{\lim }\left( x_{i,2}\left( t\right)
-v^{\left( i-1\right) }\left( t\right) \right) =o\left( \Delta
^{n-i+1}\right) ,i=1,\cdots ,n-1
\end{equation}%
where $O(\Delta ^{n-i+1})$\ denotes the approximation of $\Delta ^{n-i+1}$\
order $^{{[13]}}$\ between $x_{i,2}\left( t\right) $ and $v^{\left(
i-1\right) }\left( t\right) $. When $0<\Delta <1$, the larger the dimension $%
n$ of the differentiator is, the higher the estimation precision is.

It means that $x_{i,2}$ approximates the undelayed derivative $v^{\left(
i-1\right) }\left( t\right) $ for $i=1,\cdots ,n-1$. This concludes the
proof. $\blacksquare $

\emph{Remark 1:} Reducing peaking phenomena

In two-step differentiator (3)-(4), peaking phenomena happen due to the
infinity $\varepsilon $. In order to reduce peaking phenomena sufficiently,
we choose $\varepsilon $ as

\begin{equation}
\frac{1}{\varepsilon }=R=\left\{ 
\begin{array}{c}
R_{0}t^{p},t\leq t_{\max } \\ 
R_{0},t>t_{\max }%
\end{array}%
\right.
\end{equation}%
where $R_{0}>0$, $t_{\max }>0$ and $p\geq 1$ are chosen according to the
desired maximum error that depends on the value of $R_{\max }=R_{0}t_{\max
}^{p}$. The selection of $p\geq 1$ can make $t^{p}$ more smaller in $t\in
\left( 0,1\right) $.

\emph{Remark 2:} From (12), we know that $s^{n}\varepsilon
^{n}+s^{n-1}\varepsilon ^{n-1}k_{1}+\cdots +s\varepsilon k_{n-1}+k_{n}$, or

\begin{equation*}
s^{n}+\frac{k_{1}}{\varepsilon }s^{n-1}+\cdots +\frac{k_{n-1}}{\varepsilon
^{n-1}}s+\frac{k_{n}}{\varepsilon ^{n}}
\end{equation*}%
is a high-order integral chain, the outputs $x_{i,2}\left( t\right)
,i=1,\cdots ,n-1$, can be filtered through choosing suitable $\varepsilon $
and $k_{i}$ $\left( i=1,\cdots ,n\right) $. Moreover, the output delay
caused by high-order integral chain can be corrected by using $\Delta
+\Delta _{g}$ instead of $\Delta $ in two-step differentiator, i.e.,

\begin{eqnarray}
\dot{x}_{i,1} &=&x_{i+1,1}+\frac{k_{i}}{\varepsilon ^{i}}\left( v\left(
t-\Delta \right) -x_{1,1}\right) ,i=1,\cdots ,n-1  \notag \\
\dot{x}_{n,1} &=&\frac{k_{n}}{\varepsilon ^{n}}\left( v\left( t-\Delta
\right) -x_{1,1}\right)
\end{eqnarray}%
and

\begin{eqnarray}
\dot{x}_{i,2} &=&x_{i+1,2}+\left( \sum\limits_{j{=i}}^{{n}}\frac{1}{\left(
j-i\right) !}\frac{k_{j}}{\varepsilon ^{j}}\left( \Delta +\Delta _{g}\right)
^{j-i}\right) \left( v\left( t-\Delta \right) -x_{1,1}\right) ,  \notag \\
i &=&1,\cdots ,n-1  \notag \\
\dot{x}_{n,2} &=&\frac{k_{n}}{\varepsilon ^{n}}\left( v\left( t-\Delta
\right) -x_{1,1}\right)
\end{eqnarray}%
where $\Delta _{g}$ is the delay caused by high-order integral chain.

{\Large 4. Simulations}

In the following simulations, we select the function of $\sin (t)$\ as the
desired signal $v(t)$, and the delayed signal $v(t-\Delta )=sin\left(
t-\Delta \right) $ is taken as the measurement signal. By two-step
differentiator, the undelayed signal tracking and derivatives estimation are
obtained from the delayed signal $v(t-\Delta )=sin\left( t-\Delta \right) $.
Moreover, we know that $\frac{dv\left( t\right) }{dt}=\cos t$, and $\frac{%
d^{2}v\left( t\right) }{dt^{2}}=-\sin t$.

We select fourth-order differentiators as simulation examples.

Parameters: $\Delta =0.5s$, $k_{1}=4$, $k_{2}=6$, $k_{3}=4$, $k_{4}=1$, and

\begin{equation*}
\frac{1}{\varepsilon }=R=\left\{ 
\begin{array}{c}
100t^{7},t\leq 1 \\ 
100,t>1%
\end{array}%
\right.
\end{equation*}

4.1 Fourth-order high-gain differentiator [6, 7, 8].

\begin{eqnarray}
\dot{x}_{1} &=&x_{2}+\frac{k_{1}}{\varepsilon }\left( v\left( t-\Delta
\right) -x_{1}\right)  \notag \\
\dot{x}_{2} &=&x_{3}+\frac{k_{2}}{\varepsilon ^{2}}\left( v\left( t-\Delta
\right) -x_{1}\right)  \notag \\
\dot{x}_{3} &=&x_{4}+\frac{k_{3}}{\varepsilon ^{3}}\left( v\left( t-\Delta
\right) -x_{1}\right)  \notag \\
\dot{x}_{4} &=&\frac{k_{4}}{\varepsilon ^{4}}\left( v\left( t-\Delta \right)
-x_{1}\right)
\end{eqnarray}

Figures 1, 2 and 3 denote respectively signal tracking, estimations of
first-order and second-order derivatives by fourth-order high-gain
differentiator. $x_{1}$, $x_{2}$ and $x_{3}$ approximate respectively $%
v\left( t-\Delta \right) $, $dv\left( t-\Delta \right) /dt$ and $%
d^{2}v\left( t-\Delta \right) /dt^{2}$ but not $v\left( t\right) $, $%
dv\left( t\right) /dt$ and $d^{2}v\left( t\right) /dt^{2}$.

4.2 Two-step fourth-order differentiator. From two-step differentiator
(3)-(4), let $n=4$, we get the two-step fourth-order differentiator as:

\begin{eqnarray}
\dot{x}_{1,1} &=&x_{2,1}+\frac{k_{1}}{\varepsilon }\left( v\left( t-\Delta
\right) -x_{1,1}\right)  \notag \\
\dot{x}_{2,1} &=&x_{3,1}+\frac{k_{2}}{\varepsilon ^{2}}\left( v\left(
t-\Delta \right) -x_{1,1}\right)  \notag \\
\dot{x}_{3,1} &=&x_{4,1}+\frac{k_{3}}{\varepsilon ^{3}}\left( v\left(
t-\Delta \right) -x_{1,1}\right)  \notag \\
\dot{x}_{4,1} &=&\frac{k_{4}}{\varepsilon ^{4}}\left( v\left( t-\Delta
\right) -x_{1,1}\right)
\end{eqnarray}%
and

\begin{eqnarray}
\dot{x}_{1,2} &=&x_{2,2}+\left( \frac{k_{1}}{\varepsilon }+\frac{k_{2}}{%
\varepsilon ^{2}}\Delta +\frac{k_{3}}{2!\varepsilon ^{3}}\Delta ^{2}+\frac{%
k_{4}}{3!\varepsilon ^{4}}\Delta ^{3}\right) \left( v\left( t-\Delta \right)
-x_{1,1}\right)  \notag \\
\dot{x}_{2,2} &=&x_{3,2}+\left( \frac{k_{2}}{\varepsilon ^{2}}+\frac{k_{3}}{%
\varepsilon ^{3}}\Delta +\frac{k_{4}}{2!\varepsilon ^{4}}\Delta ^{2}\right)
\left( v\left( t-\Delta \right) -x_{1,1}\right)  \notag \\
\dot{x}_{3,2} &=&x_{4,2}+\left( \frac{k_{3}}{\varepsilon ^{3}}+\frac{k_{4}}{%
\varepsilon ^{4}}\Delta \right) \left( v\left( t-\Delta \right)
-x_{1,1}\right)  \notag \\
\dot{x}_{4,2} &=&\frac{k_{4}}{\varepsilon ^{4}}\left( v\left( t-\Delta
\right) -x_{1,1}\right)
\end{eqnarray}

Figures 4, 5 and 6 denote respectively signal tracking, estimations of
first-order and second-order derivatives by two-step fourth-order
differentiator. $x_{1,2}$, $x_{2,2}$ and $x_{3,2}$ approximate respectively $%
v\left( t\right) $, $dv\left( t\right) /dt$ and $d^{2}v\left( t\right)
/dt^{2}$ in spite of the delayed signal $v\left( t-\Delta \right) $.

4.3 Two-step fourth-order differentiator for signal with noise. We select $%
\frac{1}{\varepsilon }=6,k_{1}=4$, $k_{2}=6$, $k_{3}=4$, $k_{4}=1$.

Figures 7, 8, 9 and 10 denote respectively input signal with noise, signal
tracking, estimations of first-order and second-order derivatives by
two-step fourth-order differentiator. $x_{1,2}$, $x_{2,2}$ and $x_{3,27}$
approximate respectively $v\left( t\right) $, $dv\left( t\right) /dt$ and $%
d^{2}v\left( t\right) /dt^{2}$ in spite of the delayed signal $v\left(
t-\Delta \right) $ and noise $\delta \left( t\right) $.

From the simulations above, we find that by the normal differentiators, the
undelayed derivatives can't be obtained from the delayed signals, and time
delays exist in the outputs. By the proposed two-step linear differentiator,
we can not only obtain approximately the undelayed derivatives from delayed
signals but aslo estimate the future information from the present signals.

{\Large 5. Conclusion}

This paper develops a two-step linear differentiator to estimate the
undelayed derivatives of delayed signal. The differentiator algorithm is
composed by two step sub-differentiators. The first step sub-differentiator
estimates the derivatives at the delay, and the second step
sub-differentiator estimates the present derivatives.

\begin{figure}[h]
\begin{center}
\includegraphics[width=3.00in]{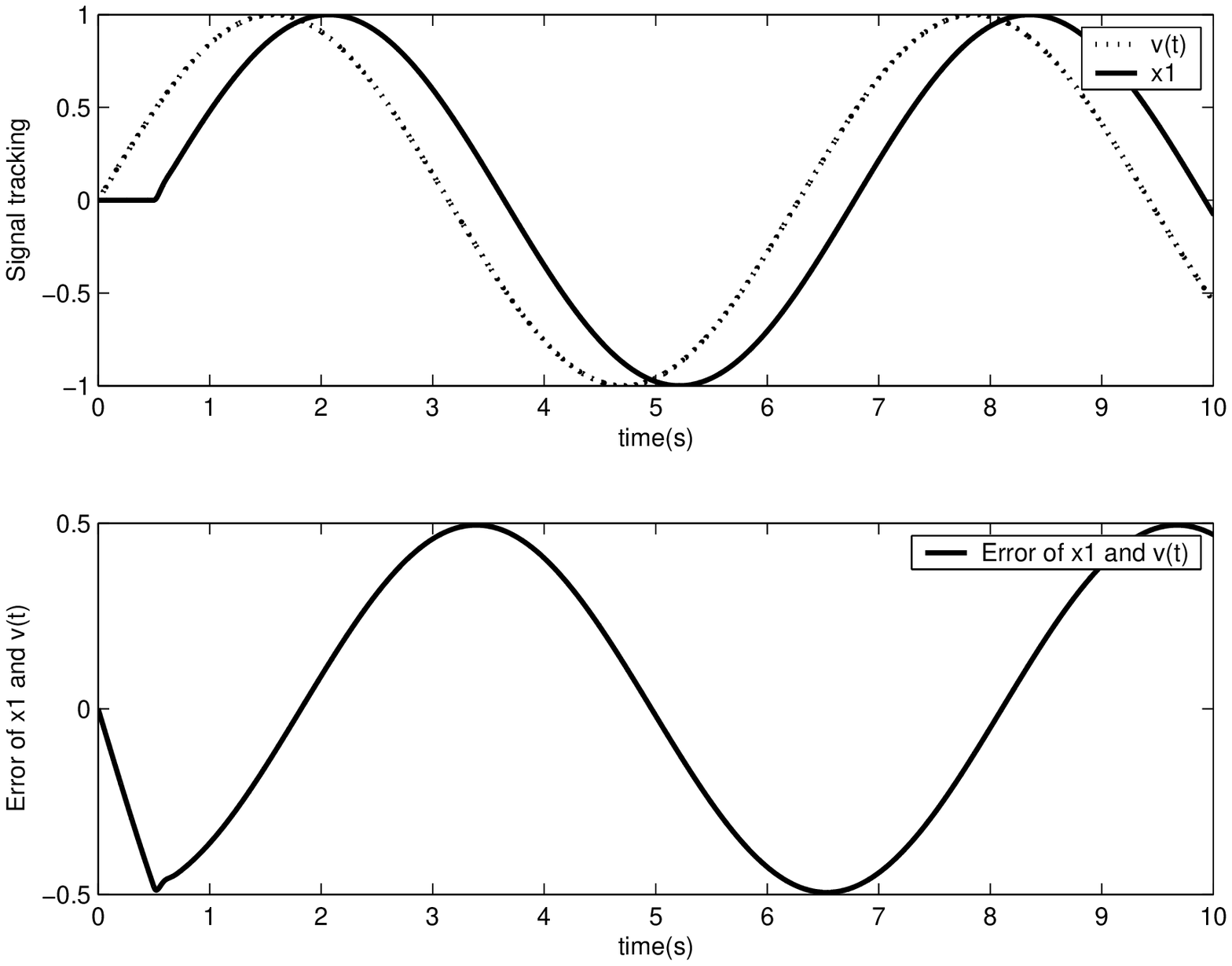}\\[0pt]
{Fig.1. Signal tracking by fourth-order high-gain differentiator (19) %
\includegraphics[width=3.00in]{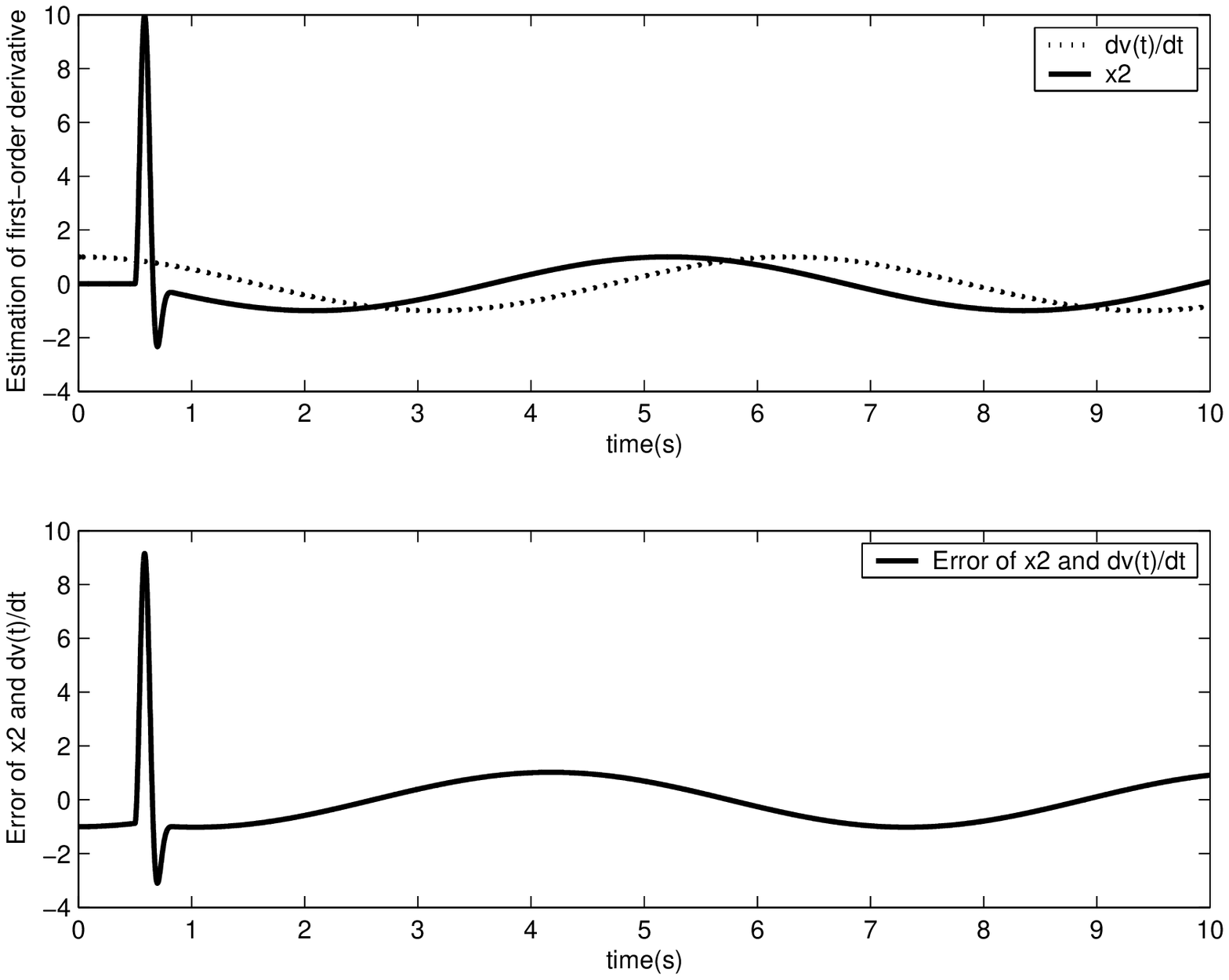}\\[0pt]
Fig.2. Estimation of first-order derivative by fourth-order high-gain
differentiator (19)\\[0pt]
\includegraphics[width=3.00in]{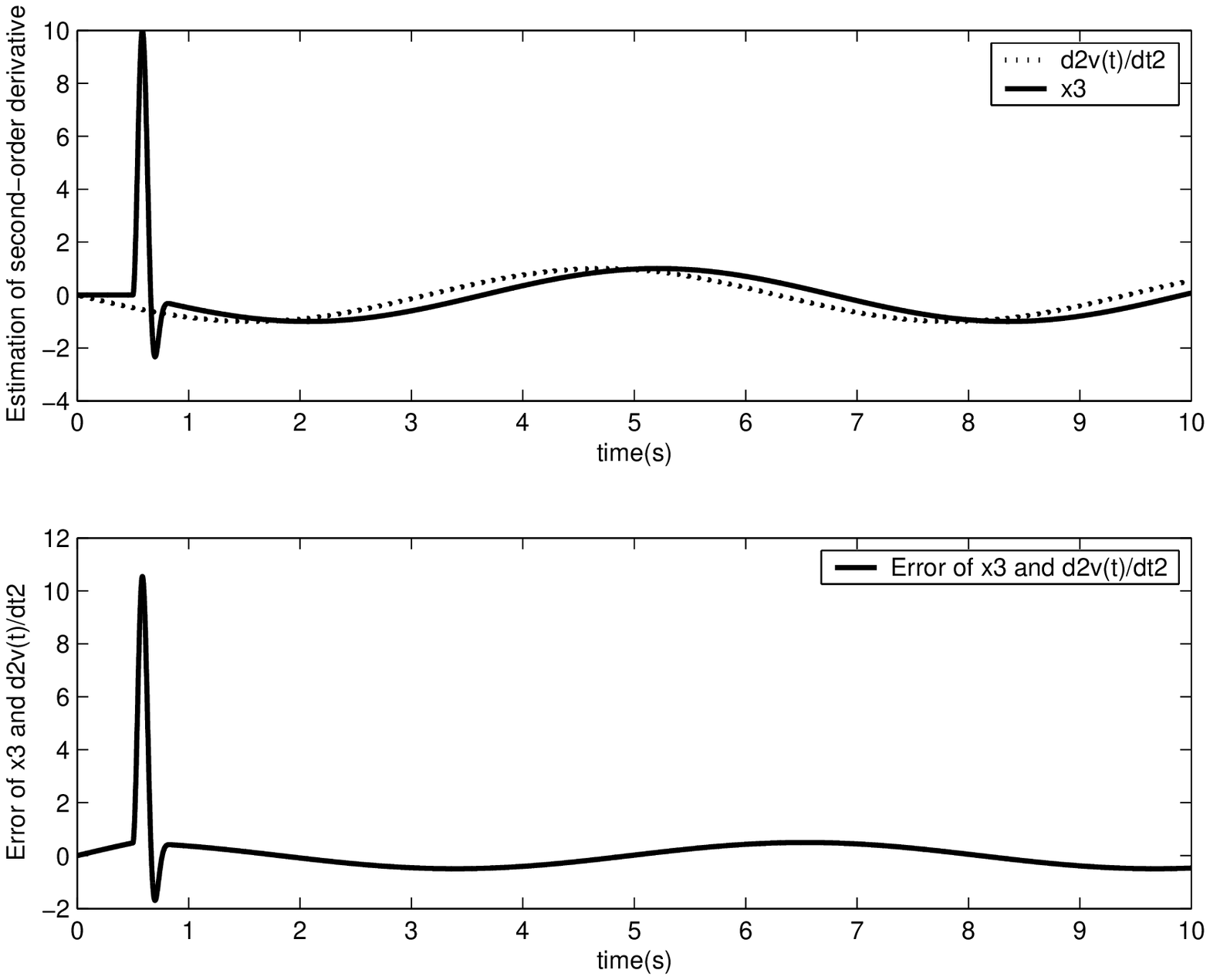}\\[0pt]
Fig.3. Estimation of second-order derivative by four-order high-gain
differentiator (19) }
\end{center}
\end{figure}

\begin{figure}[h]
\begin{center}
\includegraphics[width=3.00in]{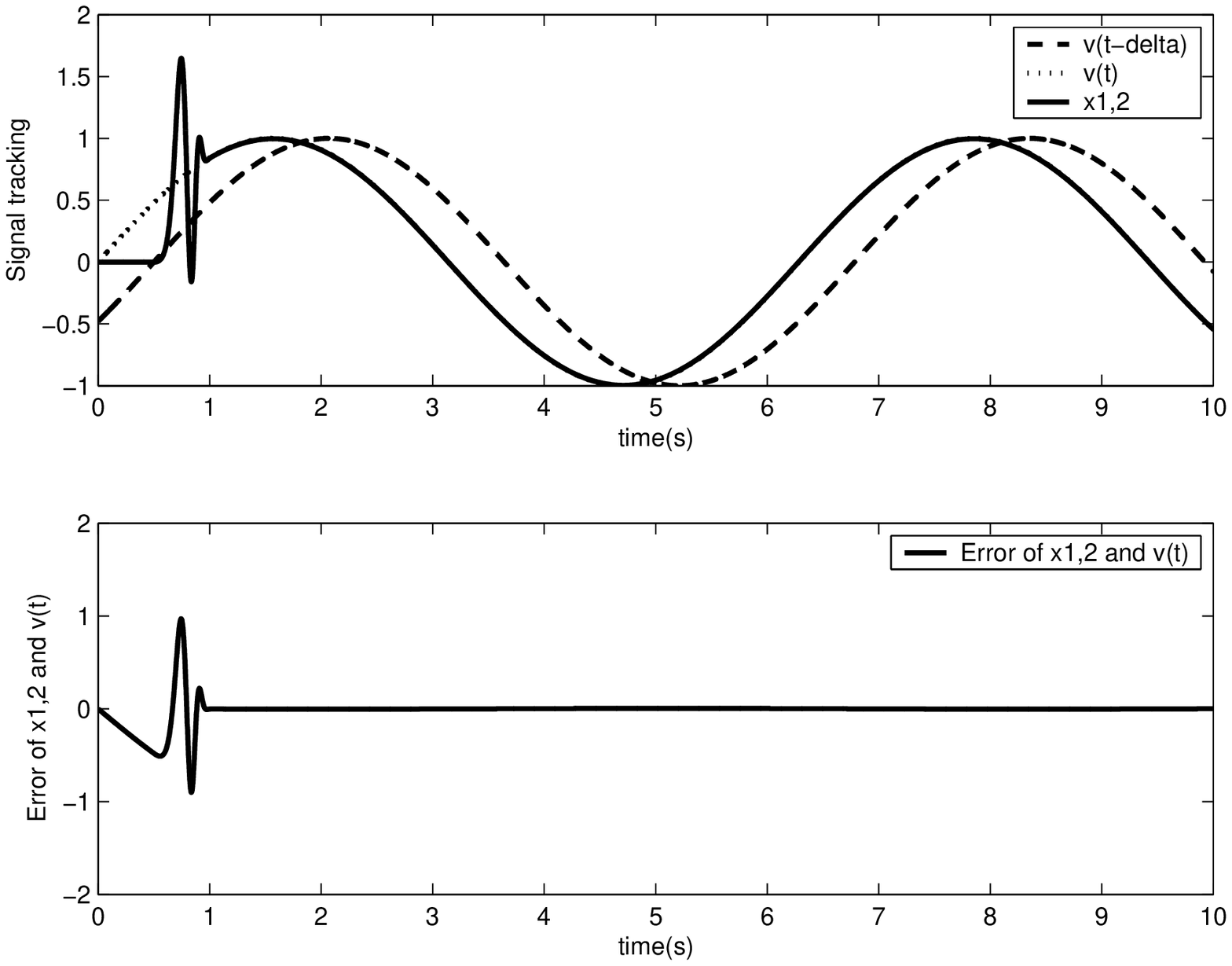}\\[0pt]
{Fig.4. Signal tracking by two-step fourth-order differentiator (20)-(21) %
\includegraphics[width=3.00in]{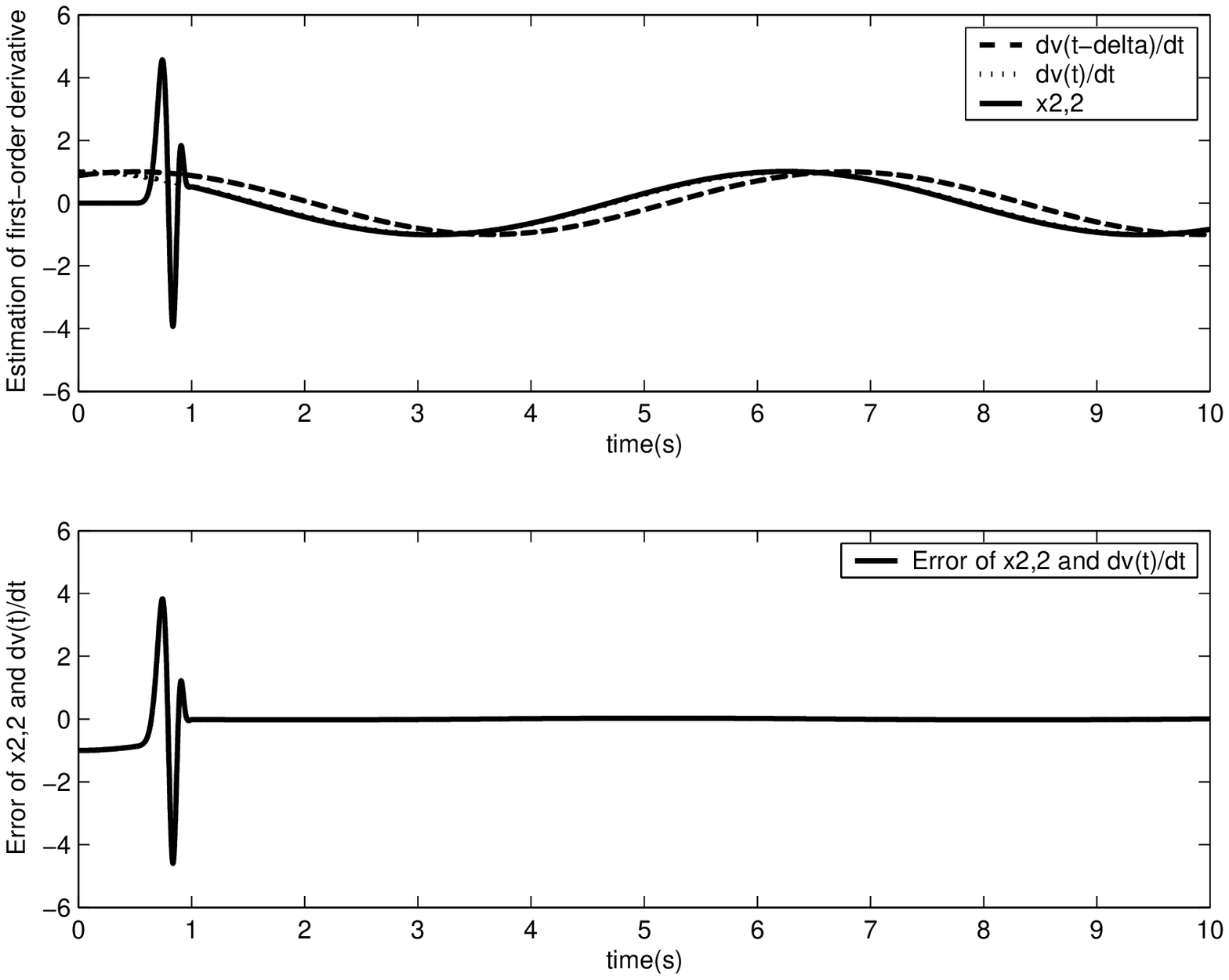}\\[0pt]
Fig.5. Estimation of first-order derivative by two-step fourth-order
differentiator (20)-(21)\\[0pt]
\includegraphics[width=3.00in]{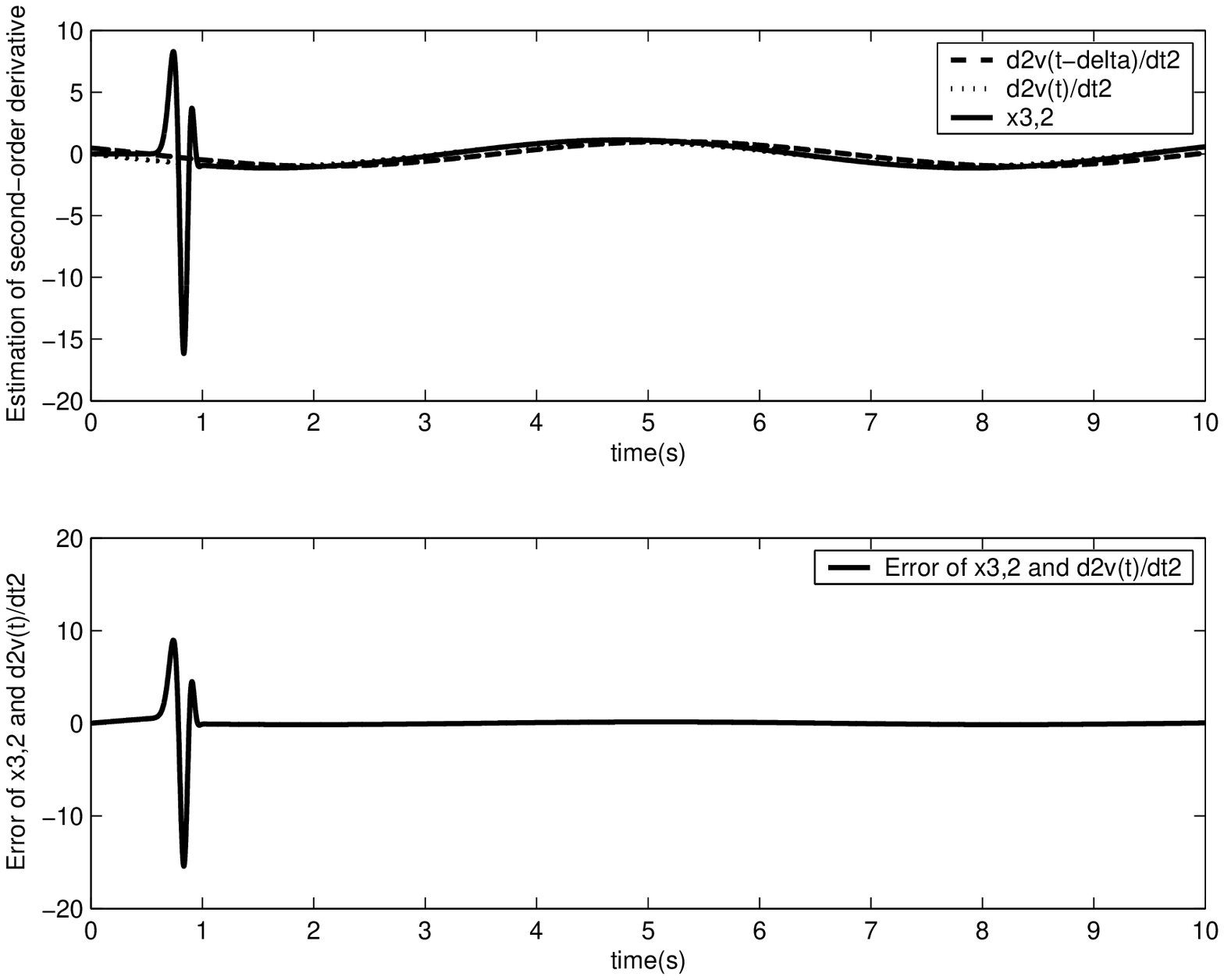}\\[0pt]
Fig.6. Estimation of second-order derivative by two-step fourth-order
differentiator (20)-(21) }
\end{center}
\end{figure}

\begin{figure}[h]
\begin{center}
\includegraphics[width=2.50in]{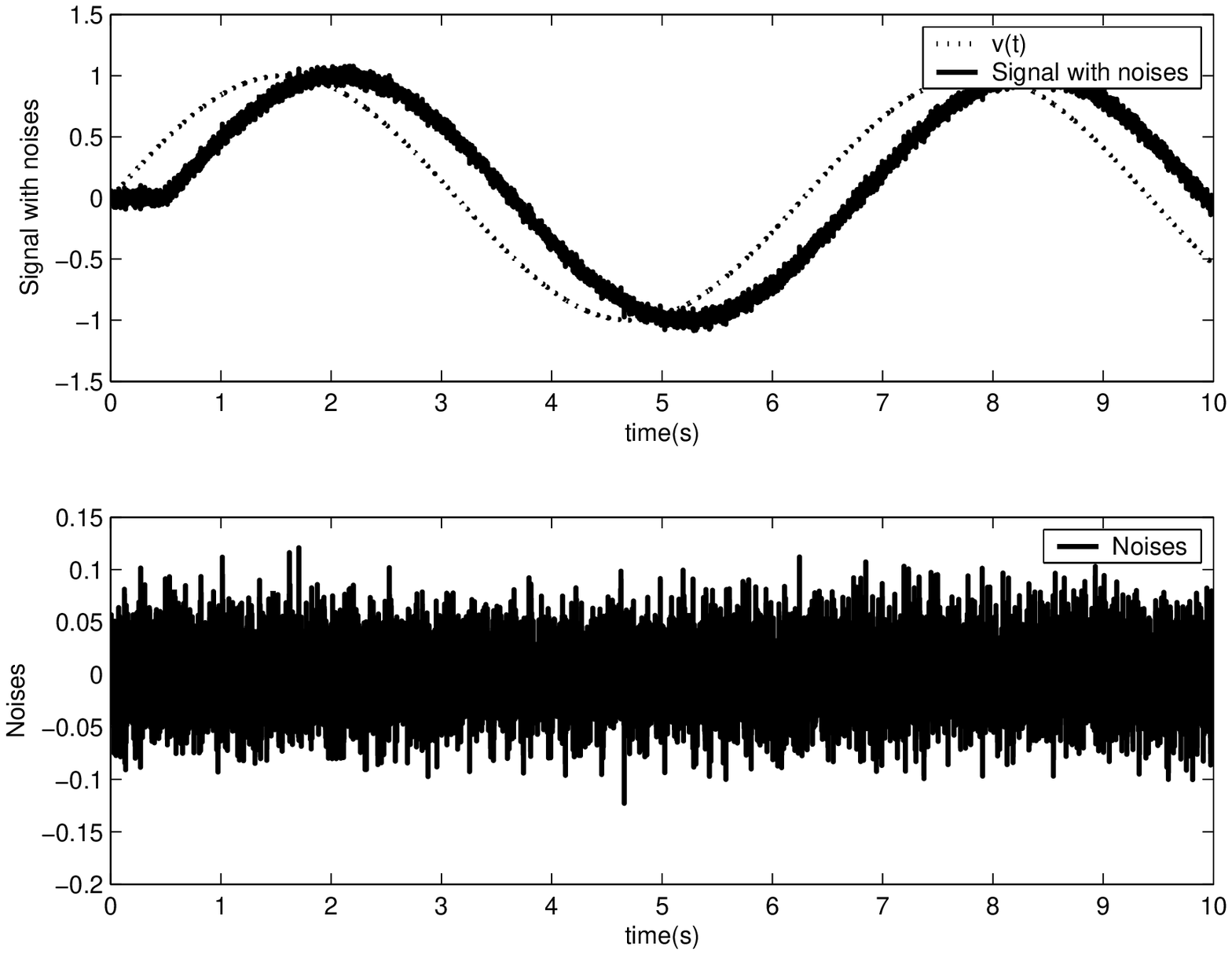}\\[0pt]
{Fig.7. Input signal with noise}\\[0pt]
\includegraphics[width=2.50in]{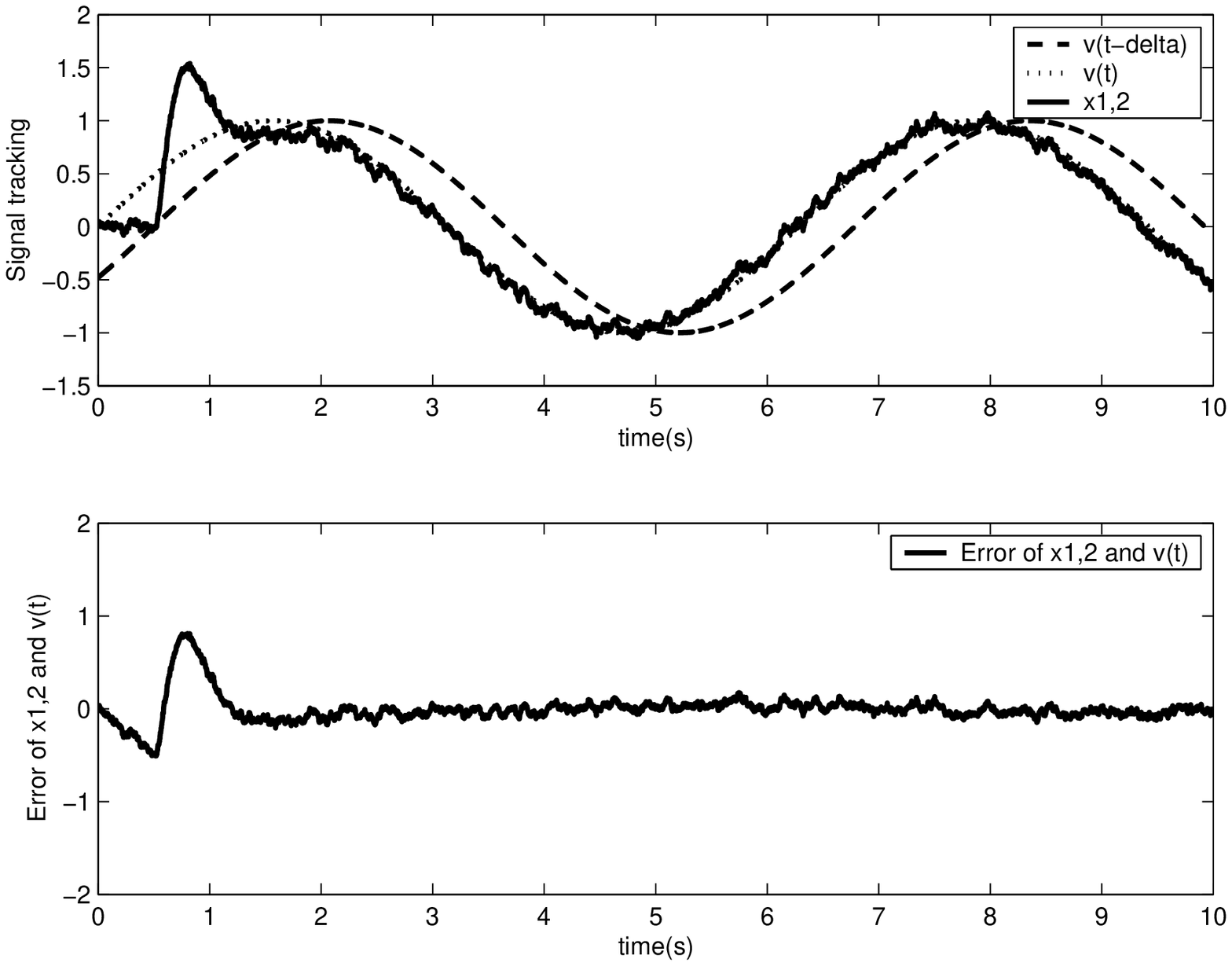}\\[0pt]
{Fig.8. Signal tracking by two-step fourth-order differentiator\\[0pt]
\includegraphics[width=2.50in]{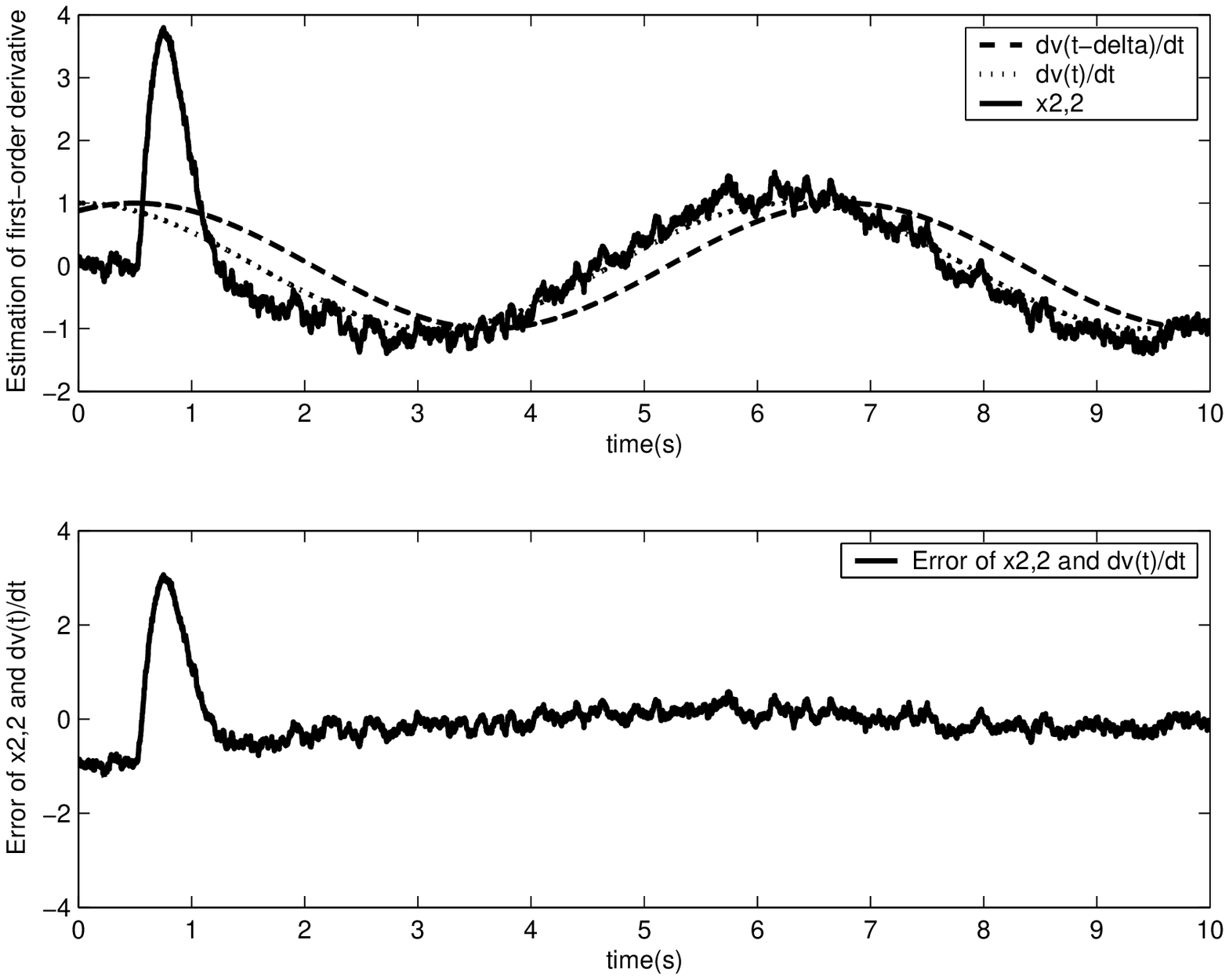}\\[0pt]
Fig.9. Estimation of first-order derivative by two-step fourth-order
differentiator\\[0pt]
\includegraphics[width=2.50in]{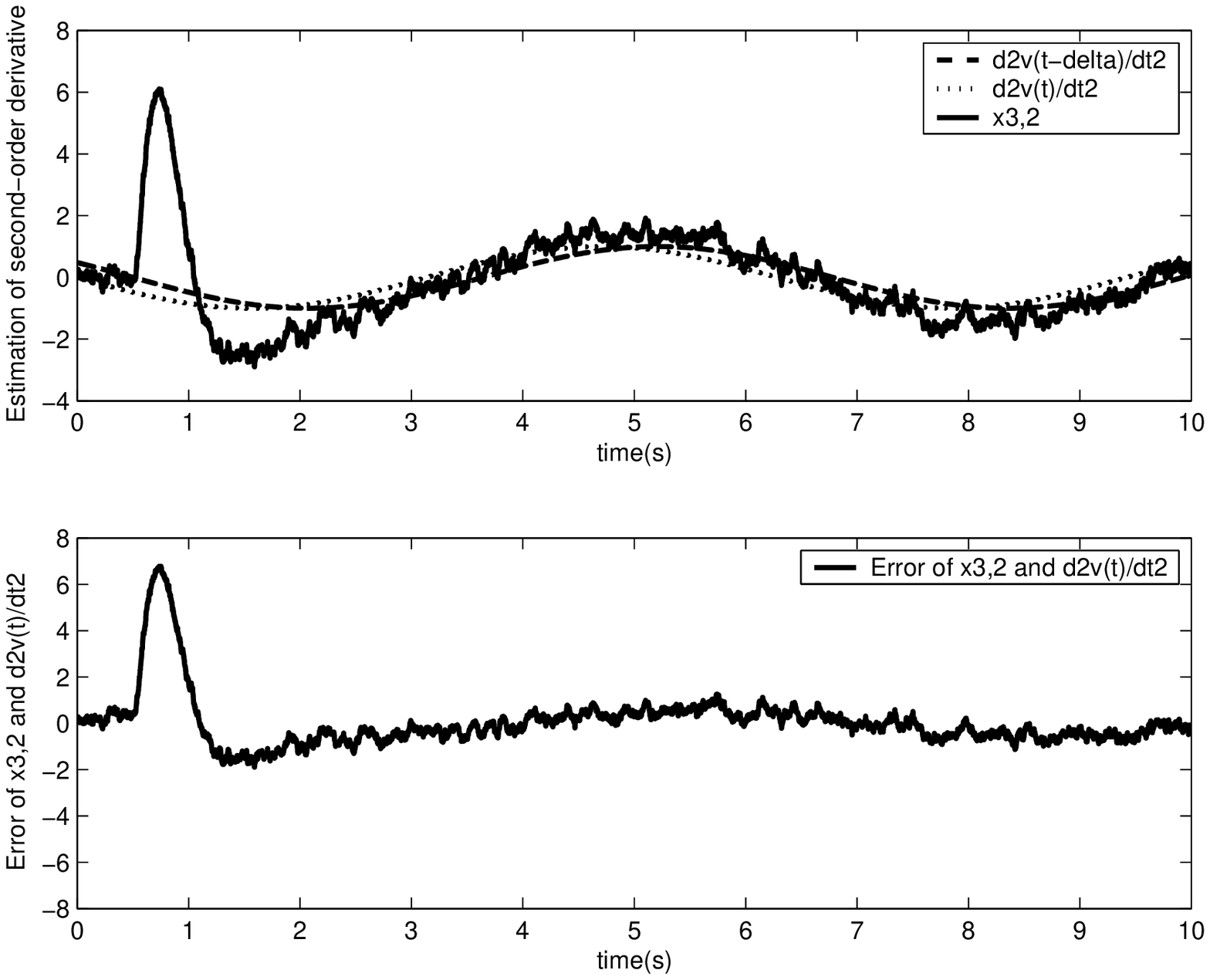}\\[0pt]
Fig.10. Estimation of second-order derivative by two-step fourth-order
differentiator }
\end{center}
\end{figure}


\begin{thebibliography}{99}
\bibitem{} J. Moffat, "The system dynamics of future warfare," European
Journal of Operational Research, 90(3), 609-618, 1996.

\bibitem{} A. Pisano and E. Usai, "Output-feedback control of an underwater
vehicle prototype by higher-order sliding modes," Automtica, 40, 1525-1531,
2004.

\bibitem{} S.-C. Pei and J.-J. Shyu, "Design of FIR Hilbert Transformers and
differentiators by eigenfilter," IEEE Trans. Acoust. Speech Signal Process.,
ASSP(37), 505-511, 1989.

\bibitem{} B. Kumar and S. C. D. Roy, "Design of digital differentiators for
low frequencies," Proc. IEEE, 76, 287-289, 1988.

\bibitem{} L. R. Rabiner and K. Steiglitz, "The design of wide-band
recursive and nonrecursive digital differentiators," IEEE Trans. Audio
Electroacoust., AU(18), 204-209, 1970.

\bibitem{} H. K. Khalil, "Robust servomechanism output feedback controller
for feedback linearizable systems," Automatica, 30, 1587-1599, 1994.

\bibitem{} A.N. Atassi, and H. K. Khalil, "Separation results for the
stabilization of nonlinear systems using different high-gain observer
designs," Systems \& Control Letters, 39, 183-191, 2000.

\bibitem{} S. Ibrir, "Linear time-derivative trackers," Automatica Vol.40,
397-405, 2004.

\bibitem{} A. Levant, "Robust exact differentiation via sliding mode
technique," Automatica, vol. 34, 379-384, 1998.

\bibitem{} A. Levant, "High-order sliding modes, differentiation and
output-feedback control," International Journal of Control, vol. 76, Nos.
9/10, 924-941, 2003.

\bibitem{} X. Wang, Z. Chen, and G. Yang, "Finite-time-convergent
differentiator based on singular perturbation technique," IEEE Transactions
on Automatic Control, Vol. 52, No. 9, 1731-1737, 2007.

\bibitem{} X. Wang, Z. Chen, and Z. Yuan, "Nonlinear tracking-differentiator
with high speed in whole course," Control Theory \&Applications, Vol. 20,
No. 6, 875-878, 2003.

\bibitem{} H. Kahlil, "Nonlinear systems," 3nd ed. Englewood Cliffs, New
Jerse: Prentice-Hall, 2002.
\end{thebibliography}
\end{document}